# Towards High-Resolution Mercury Global Topography from MESSENGER Orbital Stereo Imaging –

# A Prototype Model for the H6 Quadrangle "Kuiper"

Frank Preusker [a,*], Alexander Stark [b], Jürgen Oberst [a,b,c], Klaus-Dieter Matz [a], Klaus Gwinner [a], Tom R. Watters [d]

[a] German Aerospace Center, Institute of Planetary Research, D-12489 Berlin, Germany

[b] Technische Universität Berlin, Institute of Geodesy and Geoinformation Science, D-10623 Berlin, Germany

[c] Moscow State University for Geodesy and Cartography, RU-105064 Moscow, Russia

[d] Center for Earth and Planetary Studies, National Air and Space Museum, Smithsonian Institution, Washington, D.C. 20560-0315, USA

* Corresponding author

Tel.: +49 30 67055 446

E-mail address: frank.preusker@dlr.de (F. Preusker)

## Abstract

We selected approximately 10,500 Narrow Angle Camera (NAC) and Wide-Angle Camera (WAC) stereo images obtained by the MESSENGER spacecraft during its orbital mission with average resolutions of 150 m/pixel to compute a Digital Terrain Model (DTM) for the so-called H6 quadrangle (Kuiper), extending between 22.5°S to 22.5°N and 288.0°E to 360.0°E. Among the images, we identified about 21,100 stereo image combinations with each combination consisting of at least 3 images. We applied sparse multi-image matching to derive approximately 250,000 tie-points representing 50,000 ground points. We used the tie-points to carry out a photogrammetric block adjustment, which improves the image pointing and the accuracy of (3D) ground points from about 850 m to approximately 55 m. Subsequently, we applied high density (pixel-by-pixel) multi-image matching to derive about 45 billion tie-points. Benefitting from improved image pointing data, we compute about 6.3 billion surface points. By interpolation, we generated a DEM with a lateral spacing of 221.7 m/pixel (192 pixels per degree) and a vertical accuracy of about 30 m. The comparison of the DEM with the Mercury Laser Altimeter (MLA) tracks obtained over 4 years of MESSENGER operation reveals that the DEM is geometrically very rigid. It may be used as a reference to identify MLA outliers (when MLA operating at ranging limit) or to map offsets of laser altimeter tracks, presumably caused by residual spacecraft orbit and attitude errors. After the relevant outlier removal and corrections, MLA profiles show excellent agreement with topographic profiles from H6 with root mean square height difference of only 88 m.

## 1.    Introduction

Size, shape and surface morphology represent basic geodetic data for any planet, which includes Mercury. In March 2011, MESSENGER (MErcury Surface, Space ENvironment, GEochemistry, and Ranging) was inserted into Mercury orbit (Solomon et al., 2001, 2008) and began a comprehensive mapping mission. Several complementary techniques have been used to study Mercury's topography, which include laser altimetry (Zuber et al., 2012), measurements of radio occultation times (Perry et al., 2015), limb profiling (Oberst et al., 2011; Elgner et al., 2014), and stereo imaging (Oberst et al., 2010; Preusker et al., 2011).

While the Mercury Laser Altimeter (MLA) on MESSENGER achieves a high single-shot ranging accuracy, MESSENGER's highly eccentric orbit and northern periapsis permits adequate coverage only for areas north of the equator. Moreover, the spacing between orbital ranging tracks increases toward the equator, which reduces the areal density of the measurements. Hence, the use of laser measurements for morphological studies is limited.

The radio tracking of a spacecraft yields individual data points on the local radius of the target body when the spacecraft enters the radar shadow (ingress) or reappears (egress) (Fjeldbo et al., 1976; Perry et al., 2011, 2015). MESSENGER radio occultation measurements have been used to complement the limited MLA coverage to produce a global planetary shape model (Perry et al., 2015). Using limb imaging one may determine topographic profiles along the planetary limb which may be combined to planetary shape models (Dermott and Thomas, 1988; Thomas et al., 2007; Elgner et al., 2014). However, both radio occultation and limb profiling are hampered by the grazing viewing geometry and local variations in topography near the limb point. Moreover, the models are insufficient for morphological studies due to their limited resolutions.

While regional topographic models from stereo images have been demonstrated previously using MESSENGER data from the early flybys (Oberst et al., 2010; Preusker et al., 2011), we aim at a higher resolution global models from data of MESSENGER's orbital phase in this paper. These require the use of higher resolution images and combinations of larger numbers of images and an associated carefully designed data processing strategy to maximize data product quality within the data processing resources.

In our processing, we use best-available camera calibration data (Oberst et al., 2011). Also, as the images we use were obtained over the long period of the MESSENGER mission of four years (contrary to the flyby images), updated Mercury rotation parameters have been used to warrant correct positioning of every image in the reference frame. To demonstrate and analyze the results of the processing, we focus on a prototype DTM, covering Mercury's map quadrangle H6.

## 2.    Initial data sets and requirements

### 2.1.  The H6 Quadrangle

In order to manage the complexity and challenges of the global mapping task, we chose to derive individual regional terrain models one by one following the planetary quadrangle scheme proposed for Mercury (Greeley and Batson, 1990).

The H6 equatorial quadrangle ("Kuiper") is chosen to demonstrate production of a prototype. Like the other equatorial quadrangles, H6 extends from 22.5° S to 22.5° N latitude; H6 extends from 288.0°E to 360.0°E. Hence, it includes crater Hun Kal (0.5°S, 340°E), which defines Mercury's longitude system (Archinal et al., 2011). This affords us a verification of the correct alignment of our terrain model with the reference frame. An equatorial quadrangle was chosen, as these typically combine large numbers of WAC and NAC images, each requiring different (temperature-dependent) calibration schemes. Also, comparisons with MLA tracks are possible which extend to southern latitudes up to 16°S (see Section 6). H6 was also chosen, as it demonstrates a particular challenge of the MESSENGER data – the "hot-season gaps", see details below (Section 2.2).

The quadrangle area is also geologically rewarding, as it hosts several prominent (> 300 km) impact basins, in addition to large numbers of large craters and tectonic scarps. The H6 area was already covered by images during the flybys by Mariner 10, from which stereo topographic models had been obtained (Cook and Robinson, 2000) which may be used for comparisons (see Section 7).

## 2.2. MESSENGER orbit

The MESSENGER spacecraft was in orbit about Mercury for four years (from March 2011 to April 2015). While the spacecraft approaches the planet mainly as close as 200 km in high north polar regions areas at pericenter, the orbit apocenter is near 15,300 km above the southern hemisphere. During the spacecraft orbital period of 12 hours (later, the orbit apocenter was lowered and orbital period reduced to 8 hours), the planet rotates by approximately 3.07 degrees and consequently, consecutive orbit tracks are separated by 131 km near the equator. Moreover, gaps in the mapping scheme emerge during "hot seasons" during which the spacecraft periapsis was over the dayside of Mercury, and MDIS had limited operations for about two weeks.

## 2.3. Laser altimetry

The Mercury Laser Altimeter (MLA) performed laser pulse round-trip time of flight measurements to Mercury's surface with a repetition frequency of 8 Hz (Cavanaugh et al., 2007). Altimetry measurements are possible from ranges up to 1800 km with a single-shot ranging accuracy better than 1.0 m (Sun and Neumann, 2015). Owing to the elliptic high polar orbit, the instrument cannot reach the ground over most of the southern hemisphere. Furthermore, the attitude of the spacecraft is constrained by the orientation of the spacecraft heat shield to the Sun. Hence, for almost every instrument (among MLA) mounted on the spacecraft body was in off-nadir operation. This reduces the range distance capability and ranging accuracy of the instrument. Owing to varying spacecraft speed and ranging distance, laser footprint diameters and their spacing varied significantly (see Section 6).

## 2.4. Camera System

MESSENGER's Mercury Dual Imaging System (MDIS) consists of two framing cameras, a wide-angle camera (WAC) and a narrow-angle camera (NAC), co-aligned on a pivot platform and equipped with identical 1024x1024-pixel charge-coupled device (CCD) sensors (Hawkins et al., 2007). The WAC features 11 narrow-band filters from visible to near-infrared wavelengths and a broadband clear filter. In this paper, we used images taken by the WAC filter 7 (WAC-G), which was designed to have similar sensitivity to NAC Filter M with a maximum at 750 nm (orange).

Both cameras consist of a compact off-axis optical system that has been geometrically calibrated using laboratory as well as in-flight data (Hawkins et al., 2007, 2009). The harsh thermal environment of Mercury requires sophisticated models for calibrations of focal length and distortion of the camera. In particular, the WAC camera and NAC camera were demonstrated to show a linear increase in focal length by 0.06% to 0.10% over the typical range of temperatures (-20 – +20 °C) during operation, which causes a maximum displacement of 0.6 to 1.0 pixels. Following methods described earlier (Oberst et al., 2011) the focal length dependencies and geometric distortions for WAC and NAC were modeled using observations of star fields in different temperature regimes of the MESSENGER orbit.

### 2.5. Stereo image coverage and image selection

During the mission, MDIS acquired more than 200,000 images, most of which in the Mercury orbital phase. Owing to the eccentric orbit of the spacecraft with a pericenter in north polar regions, an imaging strategy was chosen which combined the use of the WAC-G and NAC camera to cover both hemispheres at similar resolutions.

Note that with MESSENGER's maximum orbital surface distance at apocenter, the lowest possible image resolution of NAC orbital images is about 320 m/pixel. In contrast, the highest image resolution of WAC images is about 100 m/pixel within the H6 quadrangle.

Using image footprint information, we identified all narrow-angle and wide-angle filter G images that had a resolution between 50 m to 350 m that fell into the area of the H6 quadrangle. In total we found approximately 10,500 images among them about 8,950 NAC images and about 1,550 WAC-G images, including about 150 images from the second flyby in October 2008 (where images were obtained occasionally from larger distances, Preusker et al., 2011).

The stereo-photogrammetric analyses require a favorable image- and illumination geometry, which affect quantity and quality of matched tie points (see Section 4.2) and resulting DTM points. As image coverage, image scale and illumination conditions vary substantially during the MESSENGER mapping mission, the "quality" of stereo conditions vary accordingly. We use data from our previous MESSENGER image processing (Preusker et al., 2011), to define optimal, adequate and minimal criteria for surface reconstruction (Table 1). We find "optimal stereo conditions" with at least threefold stereo coverage for about 40 percent and for "adequate stereo condition" of about 92 percent within the H6 quadrangle. For the remaining areas we have "minimal stereo conditions". In the processing we begin with areas having images at optimal stereo conditions, remaining areas are filled with images of adequate and finally minimal conditions (see Fig. 1).

To identify stereo combinations we formed a latitude/longitude grid of 0.1° x 0.1° (720 x 450 grid elements). For each grid element, we identified the images covering the area (typically: 10 – 300 images) and computed the stereo angles as well as the relevant illumination angles (i.e., sun incidence, emission, and sun phase angles) for each pair. Only those pairs were considered that had image scales differing by not more than a factor of three. All pairs were tested against the conditions of Table 1. We aim at the identification of "combinations" of images concatenated by favorable stereo conditions. Typically, we find combinations with five to eight members; combinations with less than three images were discarded from the subsequent analysis.

Several groups of stereo image combinations extending over wide areas of the grid space were identified, apparently taken at similar local time during dedicated imaging campaigns, which can be combined to "stereo networks". Two large stereo networks containing several thousands of images were identified, i.e., networks of images tied through favorable conditions, but not sharing favorable conditions with images of the other network (see Fig. 2). Several smaller stereo networks, having less than fifty images, were removed from the analysis as they contributed little new data to the area.

The first main stereo network (see Fig. 2a) consists of about 15,300 stereo combinations including 5150 images (among 4,500 NAC and 650 WAC-G images), taken at local afternoon time. The second smaller stereo network (see Fig. 2b) consists of about 5,800 stereo combinations with about 2,150 images (among 1,600 NAC and 550 WAC-G images), taken in the local morning. An inspection reveals that these networks have similar sun elevations of 20° but nearly opposite sun azimuth angles. To concatenate both networks, we created 135 threefold stereo combinations between images from the two networks manually (see Sections 4.3 and 5 for details).

All in all we found about 21,235 independent stereo combinations consisting of about 7,300 stereo images (6,100 NAC and 1,200 WAC-G images).

## 3.    Ancillary Data

We use nominal MESSENGER orbit and pointing data, as provided by the mission project as well as the alignment of the camera to the spacecraft, and geometric calibration parameters of the camera (see NAIF PDS node). All ancillary data are typically provided in the format of SPICE kernels (Acton, 1996).

Furthermore all computations are carried out in the Mercury-fixed reference frame, which was defined and agreed upon at an early stage within the MESSENGER science team (see pck00010_msgr_v23.tpc). This parameter set is a combination of libration and spin axis orientation estimates from Earth-based observations (Margot et al., 2012) and the rotation rate estimate from MESSENGER radio tracking data (Mazarico et al., 2014). While recently updated rotation parameters from co-registration of MESSENGER stereo images and laser altimeter data are available (Stark et al., 2015a) we use the above mentioned "MESSENGER reference frame" to maintain consistency with other MESSENGER data products.

By combination of laser altimetry and radio occultation data, the mean radius of Mercury was derived as 2439.4 km (Perry et al., 2015). Although Mercury exhibits a significant ellipsoidal shape with axes varying in the order of 1 km, a sphere with this radius is used as a reference for our topographic model in this paper.

## 4.  Methods for DTM generation

The construction of the DTMs follows procedures we have previously used for Mercury (Oberst et al., 2010; Preusker et al., 2011, 2015), which comprises five main tasks (Fig. 3).

### 4.1.  Pre-rectification

Based on the nominal orbit and pointing information, all 7,800 images are pre-rectified to a common map projection (here: Lambert Azimuthal projection), i.e., to one common scale, using Mercury's mean radius and a priori knowledge of the topography as reference. The processing is carried out in several stages following a pyramid strategy (see Section 4), where map scale and a priori topography for each pyramid level is selected accordingly.

This procedure is used to reduce search areas for multi-image matching (see Section 4.2). Remaining image parallaxes reflect the deviation of the a priori topography assumption to the true shape of the body or inaccurate navigation data (see Fig. 4). In order to maintain the link of this pre-rectification geometry to the raw image geometry, the raw image coordinates for each pre-rectified pixel are stored in history files.

### 4.2. Multi-image Matching

A multi-image matching technique (Wewel, 1996) is applied to the pre-rectified image data in order to derive conjugate points in each of the ~21,000 stereo combinations. The algorithm makes use of area-based correlation to derive approximate values for the image coordinates, which are refined to sub-pixel accuracy by least-squares matching. The correlation is done for each master image as the reference image with all stereo partners, i.e. all overlapping images. After the multi-image matching process, the derived image coordinates are transformed back to raw data geometry, using the history files generated during the pre-rectification. The accuracy of this back-transformation is better than one tenth of a pixel (Scholten et al., 2005).

For each stereo combination two kinds of tie-point data sets are generated. First, the images are matched in a sparse grid, usually every 20[th] pixel, whereas in a second run a higher density grid is generated. Here, every pixel of the master image is matched with its stereo partners. The lower density image point measurements (matched points) are the initial input for the bundle adjustment (see Section 4.3) whereas the high-density measurement is used for DTM generation (see Section 4.4).

Particular care was taken to match the 135 selected images between the two stereo networks with their drastically differing illumination (see Section 2.5). To enable the automatic matching, negative versions of the images from the second network were produced by inversion of pixel digital number (DN) values. These negative images were successfully matched with the images from the first network (Fig. 4).

### 4.3. Bundle Block Adjustment

The central software element is the bundle block adjustment, where we carry out a least-squares inversion of image tie-point measurements to determine the unknown six camera orientation parameters (three metric parameters for the camera position and three angular parameters for the camera pointing) for every image as well as three coordinates for each tie-point in object space. The relation between tie-point coordinates and the corresponding surface point is mathematically defined through the so-called collinearity equations (Albertz and Wiggenhagen, 2009).

Nominal navigation and reference frame data (Section 3) are used to begin the iterations. No attempt was made to further improve model parameters, e.g., planet rotation model, spacecraft orbit parameters, or camera constants, at this stage. We assume that any systematic error will be spread over the adjusted navigation data of all images and will not affect the internal rigidity of the model.

### 4.4. Object Point Calculation

The line of sight for each observation, defined by the image coordinates, the geometric calibration, and the orientation data, is computed. Lines of sight for each tie-point are combined to compute forward ray intersections using least-squares techniques. We obtain object points in Cartesian coordinates and their relative accuracies. Again, the redundancy given by multi-stereo capability allows us to accept only those object points that are defined by at least three stereo observations. Thus, we avoid occasional gross matching errors, typical for simple two-image matching.

### 4.5. DTM Interpolation

For the generation of the gridded DTM we combine the object points of all models. First, all object points are transformed from Cartesian- to spherical coordinates (latitude/longitude/radius). Height values are computed with respect to Mercury's adopted mean radius of 2439.4 km. The latitude/longitude coordinates are transformed to the standard map projection of our quadrangle (here: equidistant projection). Object points located within a DTM pixel are combined by a distance-weighted mean filtering technique (Gwinner et al., 2009), which also involves neighborhood data within a four-pixel radius. Finally for regions that lack any object point information, we applied a gap-filling algorithm using data from preceding DTM pyramid levels of reduced resolution in order to derive a raster DTM without gaps.

## 5.    Results

The five steps described above are carried out at three levels, following a pyramid strategy, where reduced map scales are chosen for pre-rectification of the images (see Section 4.1), at each step. DTMs produced in the first and second pyramid level and associated adjusted navigation data are used as a priori models for pre-rectification of images in the respective subsequent levels. We apply data snooping techniques within bundle block adjustment (see Section 4.3) to eliminate outliers (identified by large ray intersection errors) at each pyramid level. As a result of this strategy we obtain two sets of tie-point observation for each stereo combination. Here, the count of tie-points per stereo combination (depending on overlapping areas) for high-density matching ranges from about 200 thousand to 4 million tie-points and from 200 to 2,000 for coarse matching. The resulting local DTMs and rectified images are used for quality assessment.

At the final block adjustment stage, we used a set of 250,000 tie point observations from coarse matching of 7,800 images, which represent 50,000 individual ground points. Hence, the ratio between observations (545 thousand images coordinates) and unknowns (150 thousand ground points coordinates) is about 3.6. As result of the adjustment the mean accuracy of the ground points is reduced from 850 m (using nominal navigation data) to 55 m. On average, the position and the pointing of the images were adjusted by 80 m (mostly in radial direction) and 0.01°, respectively.

From our set of 21,000 stereo configurations, we derived coordinates for 6.3 billion ground points. For the gridded DTM, we set a grid spacing of 192 pixels per degree, corresponding to 221.7 m/pixel (see Fig. 5). Hence, we have 50 ground points within any DTM pixel, on average. The grid spacing was selected to reflect the mean resolution of all involved images (150 m/pixel). With the given coverage of approximately 5.9 million square kilometers and the given equirectangular projection type the H6 DTM has 8,641 lines and 13,825 samples.

The mean elevation level of the DTM is 0.88 km with respect to the reference Mercury radius, with heights ranging from -3.73 km to 5.31 km, with 95 % (2σ) of the area between -1.4 km and 2.9 km. Visual inspection reveals that the terrain is characterized by an elevated area in the east and low areas in the west. The highest point within H6 is located at (4.8N°, 358.9E°) whereas the lowest point is located at (0.4N°, 322.2E°), associated with a deep crater within the "Homer" basin.

## 6.  Comparison with laser altimetry

In the equatorial regions coverage by laser profiles is sparse (Fig. 6). Only 131 profiles (counting only those with more than 100 laser footprints) cover the area of the H6 DTM. The southernmost laser footprint is located at 16.727°S, 356.050°E (in MLASCIRDR1105020655). Ranging was performed over distances varying from 200 to 1800 km. Consequently, footprint diameters varied from 35 to 280 m (median of 117 m). The footprint-to-footprint distance, driven by ground-track velocity of the spacecraft, varied from 175 to 420 m (median of 310 m), assuming that no laser returns were missed. The ranging accuracy of MLA, depending on distance and the incidence angle of the laser pulse on the surface, varied between 0.12 (for nadir observations) and 1.5 m (Sun and Neumann, 2015).

In order to compare the H6 DTM and the MLA profiles a co-registration is performed following techniques described by Gläser et al. (2013) and Stark et al. (2015b). The optimum lateral and vertical position of the laser profile with respect to the DTM is computed by minimizing the height differences between the data sets. For most profiles, the positions can be determined with accuracy better than the size of one DTM grid element. We computed offsets of the 131 MLA profiles with respect to their nominal positions, including the associated uncertainties.

Most lateral along- and across track offsets (Fig. 7) are significantly larger than these uncertainties. The 1-σ ellipse of the corrections has principle axes of 393 and 540 m, while the center of the ellipse is offset by -143 and -45 m in cross- and along track directions, respectively. We argue that this large spreading cannot be caused by internal geometric distortions of the DTM, but is more likely to be due to errors in spacecraft orbit and attitude uncertainties or residual errors of the Mercury rotation model.

Interestingly, the radial corrections for consecutive orbits, separated by 8 hours, (Fig. 8) follow some systematic trend, with pronounced discontinuities, probably a relic from the MESSENGER orbit determination process. In fact, the discontinuities are found to match with the well-known "Orbit Segment Discontinuities" (OSD)[1] (Fig. 8), originating from the fact that orbit determination was performed in discrete time segments of one week each, with parameters of the orbit determination force model estimated separately for each week (Page et al., 2014). In one particular case, the MLA profile prior the OSD on 10.12.2014 is offset to a profile after the OSD by more than 300 m, which is much larger than the uncertainty in the co-registration or the height accuracy of the DTM. Note that the radial corrections, as opposed to the lateral corrections are not as much influenced by the uncertainty in pointing, but rather by uncertainty in the orbit. Furthermore offsets may also be due to small residual errors in the Mercury rotation model.

After co-registration, MLA and H6 DTM profiles show excellent agreement (e.g. Fig. 9). We find a height RMS of 88 m, in agreement with the estimated formal accuracy of the stereo DTM. The 1.5 km deep pit within the Catullus crater (left part of Fig. 9) as well as smaller craters and topography variations are well reproduced in the stereo DTM and in the MLA measurements. When MLA approaches its ranging limits (right part of Fig. 9), data gaps and false detections become apparent.

## 7.  Discussion and Summary

The H6 DTM was computed as a prototype, to demonstrate techniques for the joint processing of very large number of MESSENGER images. Also, in this study, we carry out a comprehensive quality assessment of our data product.

By visual inspection of craters, we estimate that the DTM has an effective lateral resolution of 5 km (see Fig. 10), in agreement with earlier estimates (Preusker et al., 2011; Stark et al., 2015b). It exceeds the nominal resolution of the global DTM (Becker et al., 2016) with a nominal pixel size of 665 m. Futher, it greatly exceeds the resolution of previous Mariner-10-based DTM products (Cook and Robinson, 2010). Based on this comparison we can conclude that the DTM obtained in this work provides the best currently available topographic data set for the H6 Mercury quadrangle. Topographic landforms like crater rims, pits, ejecta rays, and thrust faults are precisely reconstructed in the DTM and are suitable for a detailed geological analysis.

---

[1] See comments in http://naif.jpl.nasa.gov/pub/naif/pds/data/mess-e_v_h-spice-6-v1.0/messsp_1000/data/spk/msgr_040803_150430_150430_od431sc_2.bsp

The model features three well known impact basins Sanai (490 km), Homer (314 km), and Renoir (246 km) as listed in the catalog by Fasset et al., (2012), as well as an unnamed basin west of Homer (about 400 km diameter), for which detailed studies of morphology can be made (see Fig. 5). In addition, the model shows a multitude of craters (we estimate more than 50 craters larger than 100 km), for which reliable measurements of depth-to-diameter ratios are possible (Fig. 11).

Also found in the higher elevation intercrater plains of the Kuiper quadrangle are lobate scarps, the surface expression of thrust faults formed as a result of global contraction (Watters et al., 2015; Solomon et al., 2008; Byrne et al., 2014). The most prominent is Santa Maria Rupes (~6°N, 340°E), a scarp with about 500 m of relief. Accurate topography of lobate scarps is critical to constraining models of the geometry and depth of the thrust faults associated with lobate scarps (Watters et al., 2016; Egea-Gonzalez et al., 2012). Estimates of the maximum fault depth can be used infer the mechanical and thermal structure of Mercury's lithosphere (Watters and Nimmo, 2010; Egea-Gonzalez et al., 2012).

Furthermore, the DTM can be used as a precious tool for the ortho-rectification and co-registration of MDIS images at resolutions of better than 200 m/pixel (Fig. 12). Within the processed images are two high-resolution images (50 m/pixel) and three images with 130 m/pixel, which contain the small crater "Hun Kal", which defines Mercury's longitude system. While the nominal location of "Hun Kal" is at 0.5°S and 340°E, we measure 0.465°S 339.995°E (see Table 2), attesting to the correct positioning of the DEM within approx. 212 m, i.e., approximately one DTM pixel (Fig. 12). While it is conceivable to shift the DTM to match the position of Hun Kal, no such effort is made, in order to maintain consistency of our DTM with other MESSENGER data products. Hence, the correct positioning of the DTM in Mercury's longitude and prime meridian system remains to be accomplished.

The DTM is found to be geometrically very rigid with minimal geometric distortions when compared with MLA data. Here, the offsets which are observed between MLA profiles and the stereo DTM are mostly due to errors in the orbit and attitude of the spacecraft during acquisition of the MLA profiles, or possibly due to residual errors of the Mercury rotation model. Hence, the H6 DTM can be used as a reference to test the positioning and pointing of the MLA tracks or to determine unknown parameters of Mercury rotation, as was previously demonstrated on earlier DTM models (Stark et al., 2015a).

Considering the successful processing scheme developed in this paper, we foresee the production of DTMs for the other Mercury quadrangles and accomplishment of a high-resolution global DTM for the entire planet. This will allow us more complete modeling of the MESSENGER orbit and Mercury rotation as well as comprehensive studies of Mercury's global morphology.

**Acknowledgements**

The DTM is available at the Planetary Data System (PDS downlink). Alexander Stark was funded by a grant from the German Research Foundation (OB124/11-1). Jürgen Oberst greatly acknowledges being hosted by MIIGAiK and supported by the Russian Science Foundation under Project 14-22-00197.

Furthermore we like to thank MESSENGER's MLA and MDIS team members for their support, especially Gregory Neumann, Erwan Mazarico, Haje Korth, Mike Reid and Susan Ensor.

**References**:

Acton, C. H., 1996. Ancillary data services of NASA's Navigation and Ancillary Information Facility. Planet. Space Sci. 44**,** 65-70.

Albertz, J., Wiggenhagen, M. 2009, Guide for Photogrammetry and Remote Sensing. Wichmann Verlag, Heidelberg, Germany.

Archinal, B.A., et al., 2011. Report of the IAU Working Group on Cartographic Coordinates and Rotational Elements: 2009. Celest. Mech. Dyn. Astron. 109, 101–135. doi:10.1007/s10569-010-9320-4.

Becker K.J., et al., 2016, First global digital elevation model of Mercury, 47[th] Lunar and Planetary Science Congress, abstract 2959, http://astrogeology.usgs.gov/search/map/Mercury/Topography/MESSENGER/Mercury_Messenger_USGS_DEM_Global_665m

Byrne, P. K., et al., 2014. Mercury's global contraction much greater than earlier estimates. Nature Geoscience. 7, 301–307.

Cavanaugh, J.F., et al., 2007. The Mercury Laser Altimeter instrument for the MESSENGER mission. Space Sci. Rev. 131, 451–479. doi:10.1007/s11214-007-9273-4.

Cook, A.C., Robinson, M.S., 2000. Mariner 10 stereo image coverage of Mercury. J. Geophys. Res. 105, 9429–9443.

Dermott, S. F., Thomas, P. C., 1988. The shape and internal structure of Mimas. Icarus. 73**,** 25-65.

Egea-Gonzalez, I., et al., 2012, Depth of faulting and ancient heat flows in the Kuiper region of Mercury from lobate scarp topography, Planet. Space Sci., 60,193–198.

Elgner, S., et al., 2014. Mercury's global shape and topography from MESSENGER limb images. Planet. Space Sci. 103, 299–308. doi:10.1016/j.pss.2014.07.019.

Fassett, C. I., et al. 2012, Large impact basins on Mercury: Global distribution, characteristics, and modification history from MESSENGER orbital data, J. Geophys. Res, E00L08.

Fjeldbo, G., et al., 1976. The occultation of Mariner 10 by Mercury. Icarus 29**,** 439-444.

Gläser, P., et al., 2013. Co-registration of laser altimeter tracks with digital terrain models and applications in planetary science. Planet. Space Sci. 89, 111-117.

Greeley, R., Batson, G., 1990, Planetary Mapping, Cambridge University Press.

Gwinner, K., et al., 2009. Derivation and validation of high-resolution digital terrain models from Mars Express HRSC-data. Photogramm. Eng. Remote Sens. 75, 1127–1142.

Hawkins III, S.E., et al.,  2007. The Mercury Dual Imaging System on the MESSENGER spacecraft. Space Sci. Rev. 131, 247–338. doi:10.1007/s11214-007-9266-3.

Hawkins III, S.E., et al., 2009. In-flight performance of MESSENGER's Mercury Dual Imaging System. In: Hoover, R.B., Levin, G.V., Rozanov, A.Y., Retherford, K.D. (Eds.), Instruments and Methods for Astrobiology and Planetary Missions. SPIE Proceedings, vol. 7441, paper 7441A-3, 12 pp., SPIE, Bellingham, Wash.

Margot, J.L., et al., 2012. Mercury's moment of inertia from spin and gravity data. J. Geophys. Res. 117, E00L09. doi:10.1029/2012JE004161.

Mazarico, E., et al., 2014. The gravity field, orientation, and ephemeris of Mercury from MESSENGER observations after three years in orbit. J. Geophys. Res., 119, 12.

Oberst, J., et al.., 2010. The morphology of Mercury's Caloris basin as seen in MESSENGER stereo topographic models. Icarus 209, 230–238.

Oberst, J., et al., 2011. Radius and limb topography of Mercury obtained from images acquired during the MESSENGER flybys, Planet. Space Sci. 59, 1918-1924.

Page, B. R., et al., 2014. Tuning the MESSENGER State Estimation Filter for Controlled Descent to Mercury Impact. AIAA/AAS Astrodynamics Specialist Conference. American Institute of Aeronautics and Astronautics.

Perry, M. E., et al., 2011. Measurement of the radius of Mercury by radio occultation during the MESSENGER flybys. Planet. Space Sci., 59, 1925-1931.

Perry, M. E., et al., 2015. The low-degree shape of Mercury. Geophys. Res. Lett. 42, 6951-6958.

Preusker, F., et al., 2011, Stereo topographic models of Mercury after three MESSENGER flybys, Planet. Space Sci., 59, 1910–1917

Preusker, F., et al., 2015, Shape model, reference system definition, and cartographic mapping standards for comet 67P/Churyumov-Gerasimenko – Stereo-photogrammetric analysis of Rosetta/OSIRIS image data, A&A, 583, A33

Scholten, F., et al., 2005. Mars Express HRSC data processing - methods and operational aspects. Photogramm. Eng. Remote Sens. 71, 1143–1152.

Solomon, S. C., et al., 2001. The MESSENGER mission to Mercury: scientific objectives and implementation. Planet. Space Sci. 49, 1445 - 1465.

Solomon, S.C., et al., 2008. Return to Mercury: a global perspective on MESSENGER's first Mercury flyby. Science 321, 59–62.

Stark, A., et al., 2015a. First MESSENGER orbital observations of Mercury's librations. Geophys. Res. Lett. 42, 7881-7889

Stark, A., et al., 2015b. Mercury's rotational parameters from MESSENGER image and laser altimeter data: A feasibility study. Planet. Space Sci. 117, 64-72.

Sun, X., Neumann, G. A., 2015. Calibration of the Mercury Laser Altimeter on the MESSENGER spacecraft. IEEE Transactions on Geoscience and Remote Sensing. 53, 2860-2874.

Thomas, P. C., et al., 2007. Shapes of the Saturnian icy satellites and their significance. Icarus. 190, 573-584.

Watters, T.R., Nimmo, F., 2010, The Tectonics of Mercury, Planetary Tectonics, Cambridge Univ. Press, New York, p. 15-80.

Watters, T.R., et al., 2015, Distribution of large-scale contractional tectonic landforms on Mercury: Implications for the origin of global stresses, Geophys. Res. Lett., 42.

Watters, T.R., et al., 2016, Fault-Bound Valley Associated with the Rembrandt Basin on Mercury, submitted to Geophys. Res. Lett.

Wewel, F., 1996. Determination of conjugate points of stereoscopic three line scanner data of Mars 96 mission. Int. Arch. Photogram. Remote Sensing 31, 936–939.

Zuber, M.T., et al., 2012. Topography of the northern hemisphere of Mercury from MESSENGER laser altimetry. Science 336, 217–220. doi:10.1126/science.1218805.

## Tables

**Table 1**

Optimal, adequate and minimal parameter ranges for key observing attributes are given for stereo-photogrammetric analyses. The lower limit of the stereo angles is specified with two values indicating whether one pair to another pair within the set can have a lower minimum angle.

|  | Optimal | Adequate | Minimal |
|---|---|---|---|
| *Parameter (degrees)* | | | |
| Illumination variation | 0 - 10 | 0 - 10 | 0 - 10 |
| Stereo angle | 15,15 - 65 | 5,15 - 65 | 5,12 - 75 |
| Incidence angle | 5 - 55 | 5 - 80 | 5 - 90 |
| Emission angle | 0 - 55 | 0 - 65 | 0 - 70 |
| Sun phase angle | 5 - 180 | 5 - 180 | 5 - 180 |
| *Stereo Coverage - Number of views (percent)* | | | |
| 3 | 23.5 | 12.8 | 8.1 |
| 4 | 11.2 | 18.9 | 13.9 |
| 5 | 4.0 | 19.4 | 19.7 |
| >5 | 1.5 | 40.5 | 58.3 |
| Total | 40.2 | 91.6 | 100.0 |

**Table 2**

Measurements of latitude and longitude position of Hun Kal crater covered by five images.

| Image name | Latitude [degrees] | Longitude [degrees] | Image scale [m/pixel] |
|---|---|---|---|
| EN0131770954M | -0.463 | 339.996 | 126 |
| EN1002521325M | -0.465 | 339.996 | 54 |
| EN1004678277M | -0.466 | 339.994 | 134 |
| EN1004678315M | -0.465 | 339.993 | 134 |
| EN1005053163M | -0.467 | 339.995 | 47 |
| Average | -0.465 | 339.995 | |

# Figures

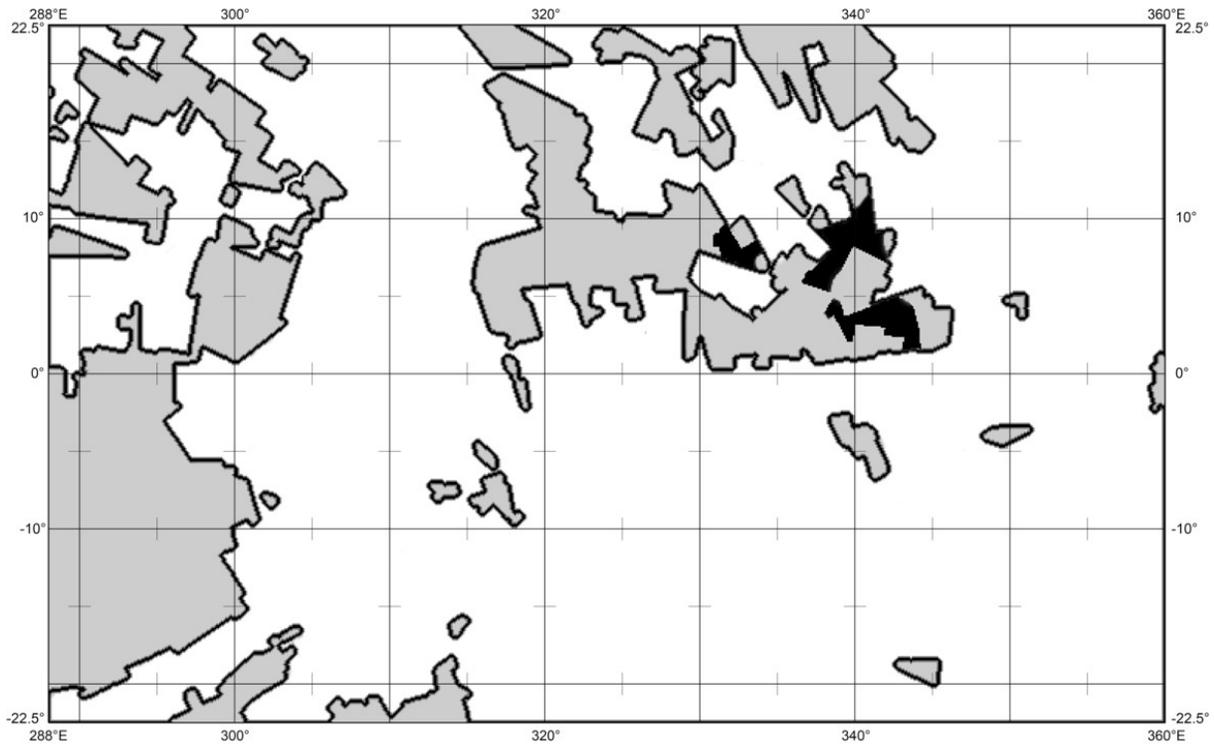

**Fig. 1.** Stereo coverage within the H6 quadrangle. Areas with "optimal" stereo coverage are marked in white. Areas in gray have "adequate" stereo coverage, areas in black have "minimal" stereo coverage (see text for further details).

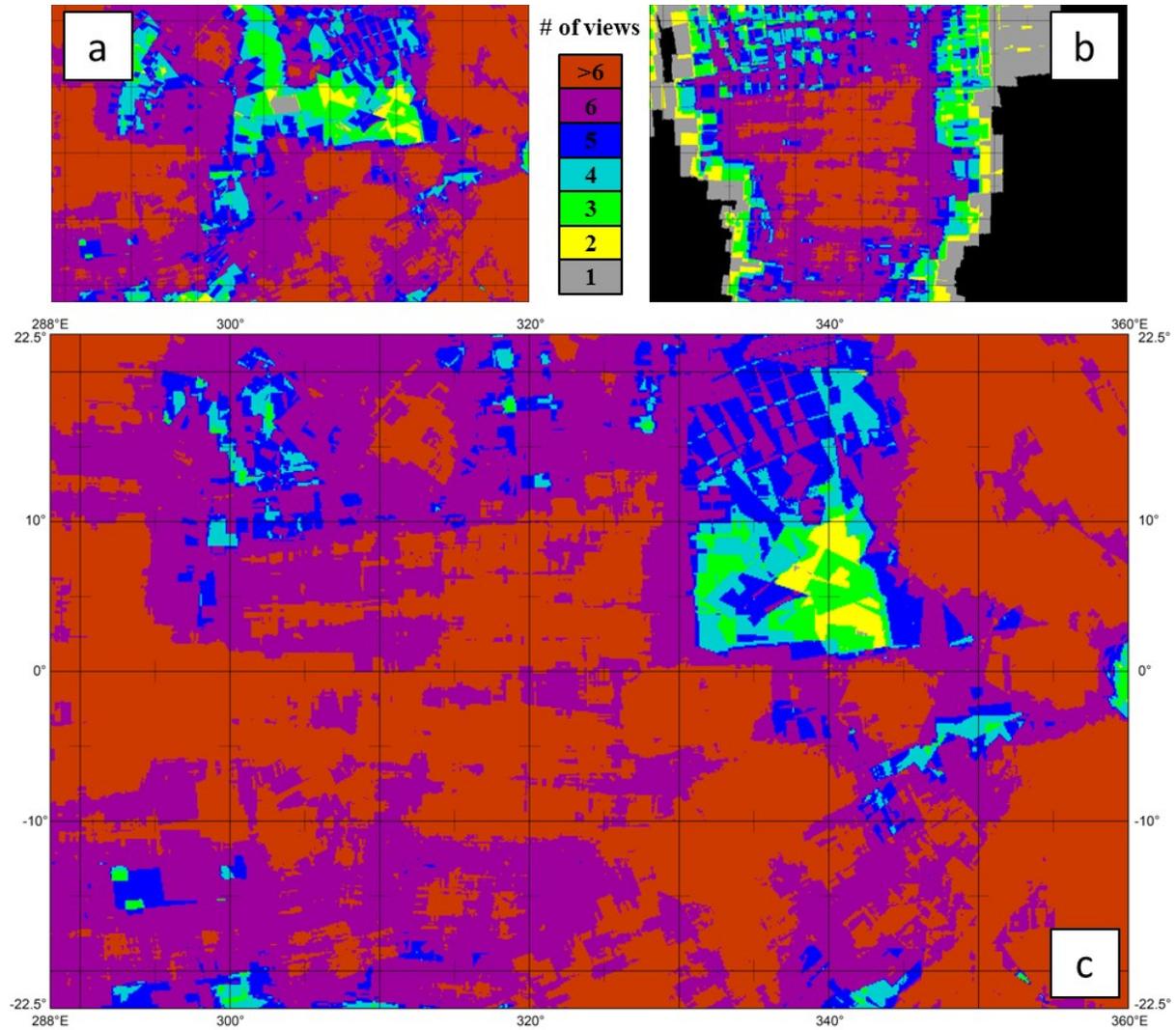

**Fig. 2**. Stereo coverage map of H6 quadrangle with "adequate" stereo conditions (see Table 1). The different colors indicate the number of stereo pairs. (a) and (b) show the coverage of network 1 and 2 (see Section 2.5) representing images taken in two different illumination regimes. (c) shows the final coverage of H6 quadrangle after merging of the two networks.

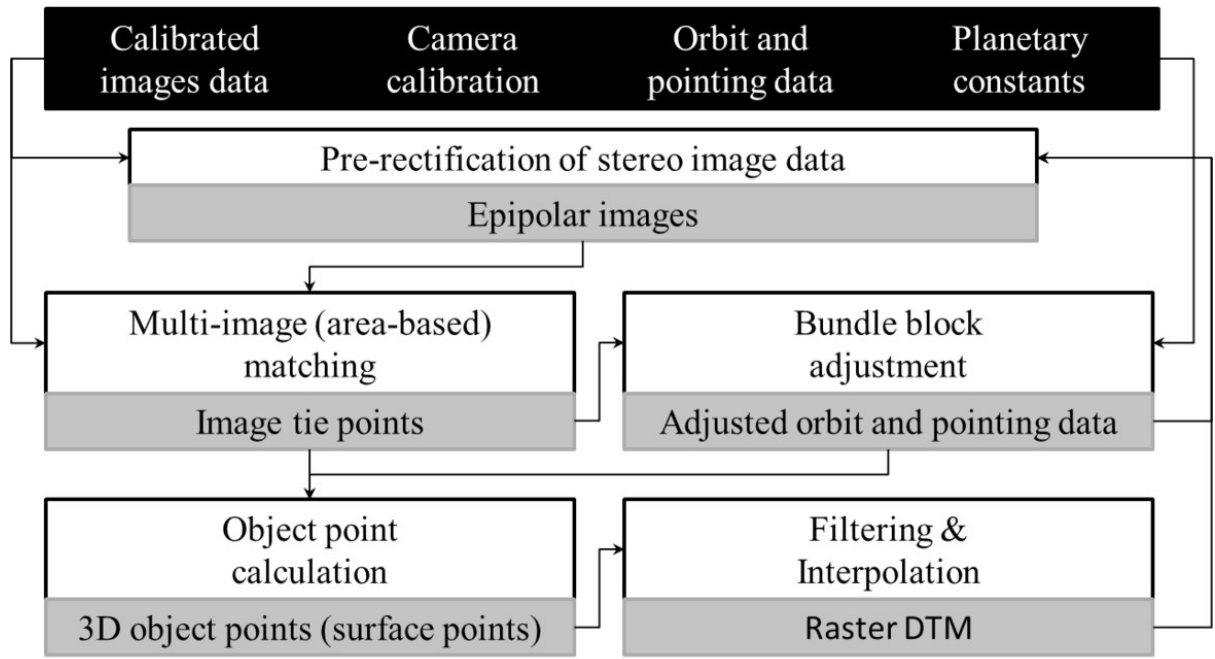

**Fig. 3.** Block diagram of the data processing sequence. Black boxes indicate input data, which enter the main five processing tasks described in Section 4. Gray boxes indicate resulting data.

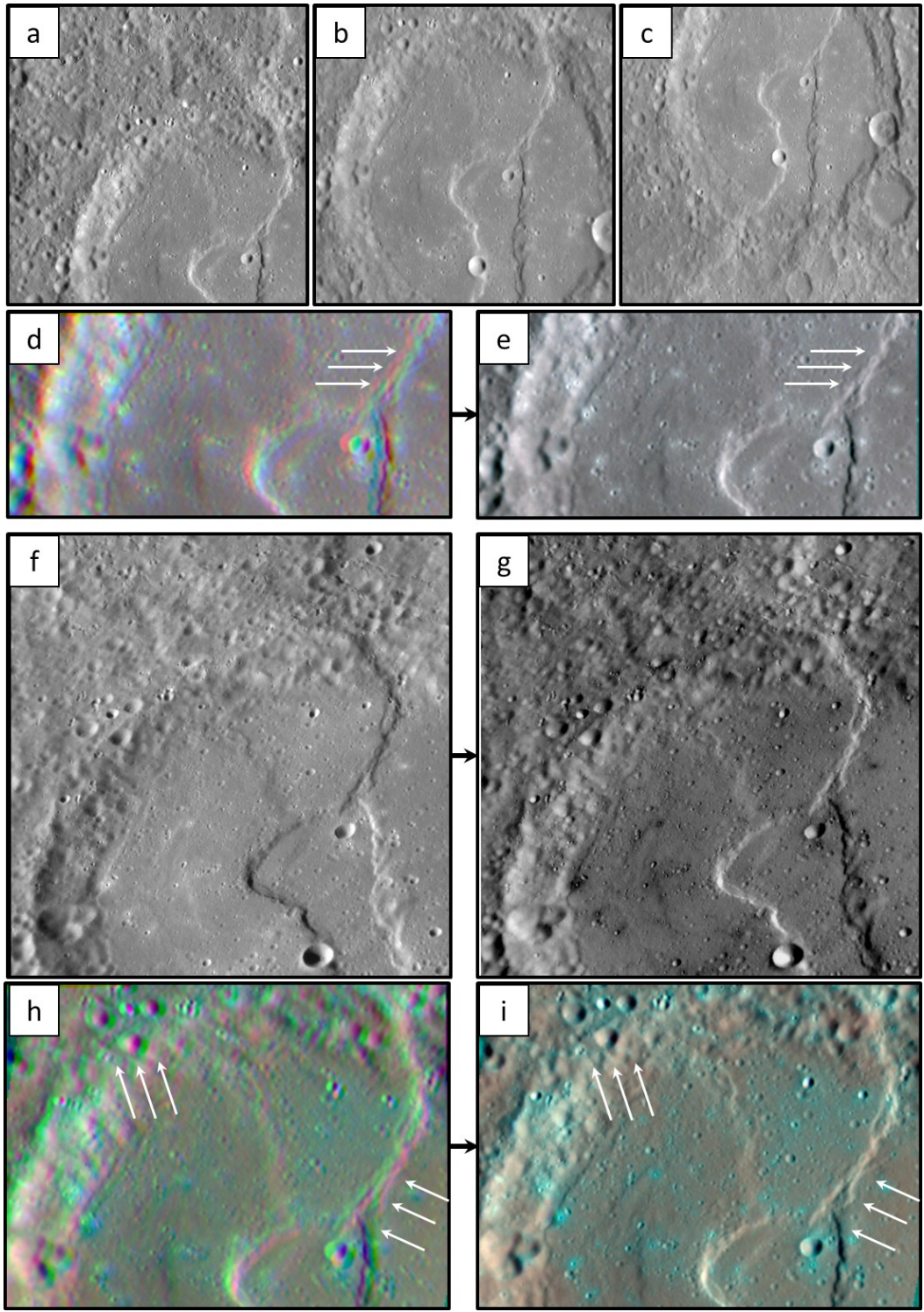

**Fig. 4.** Three selected NAC images (EN0131770853M, EN104823894M, and EN1063526637M) of a stereo configuration from stereo network 1 covering Tarkur crater (top row, a-c) and associated processed images, which were rectified using the sphere and nominal orientation (d) and using the H6 DTM and adjusted orientation (e) respectively. (f) shows a NAC image (EN0228804649M) from network 2, which is grey-scale inverted (g) and combined with stereo partners (b and c) to connect both main stereo networks (see Section 2.5). (h) and (g) show once again the associated processed images. Note the much improved co-registration of the 3 images of each combination indicated by white arrows.

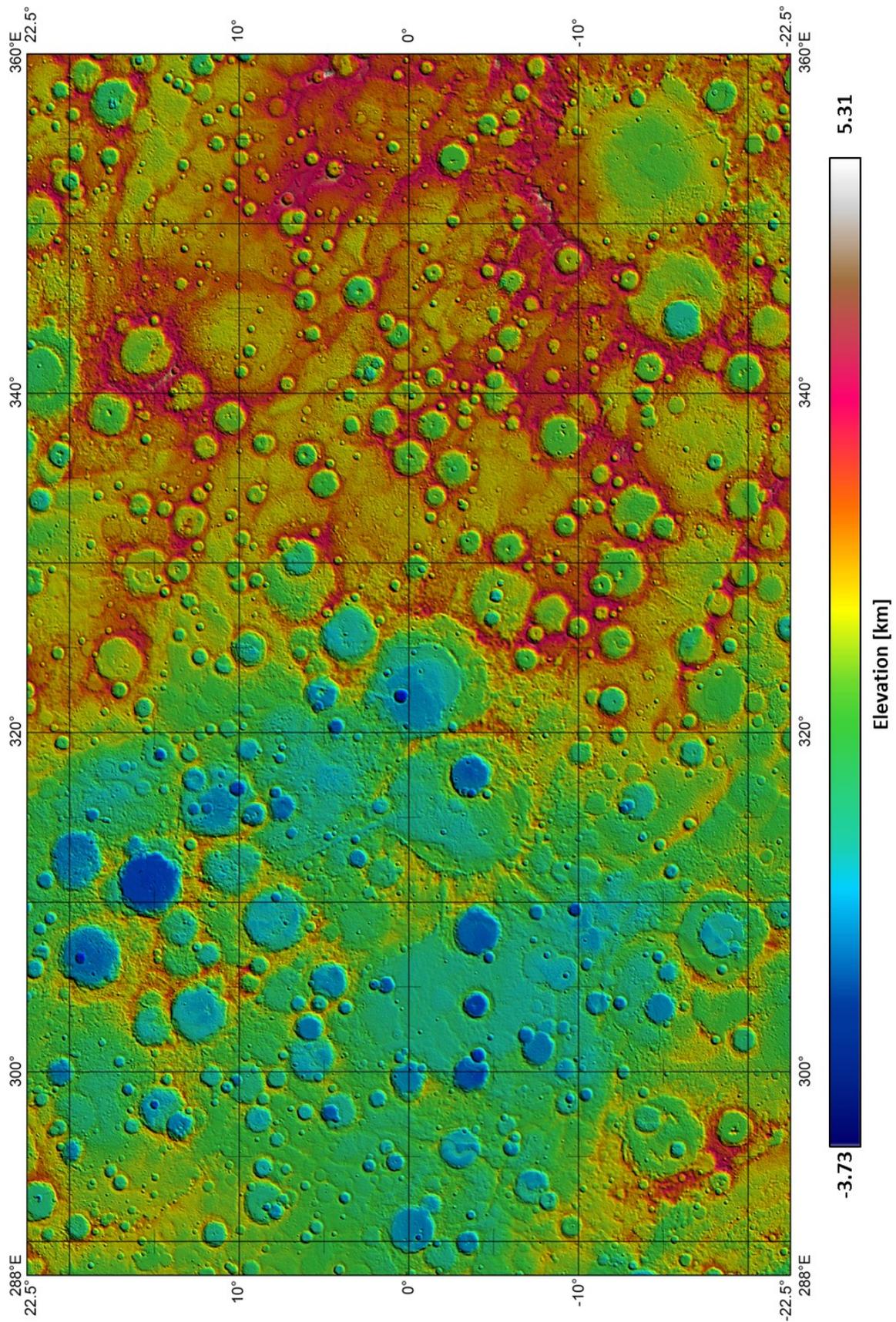

**Fig. 5.** Hill-shaded color-coded H6 DTM in equidistant projection with lateral grid spacing of 192 pixels per degree (~222 meters per pixel). Elevations are given w.r.t. Mercury's reference sphere of 2439.4 km radius.

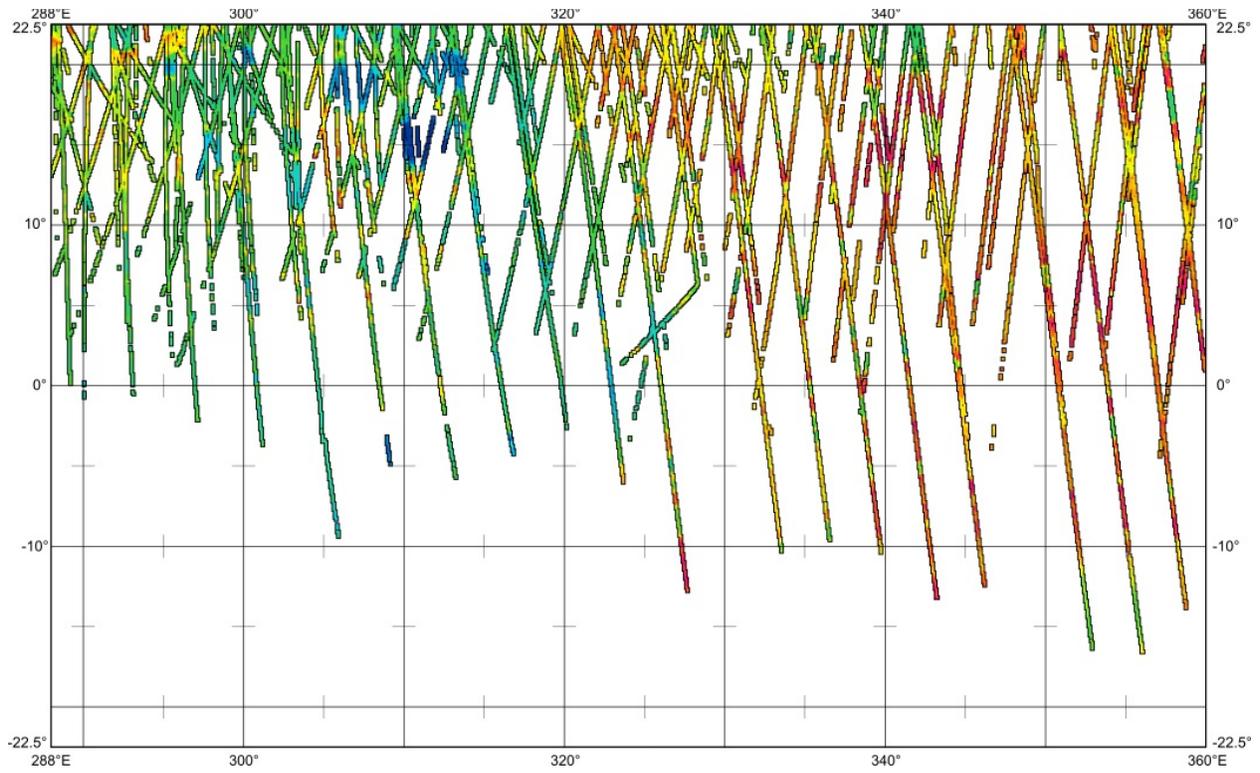

**Fig. 6.** MLA profiles within the same area as the H6 DTM. MLA heights are color-coded using the color bar as in Fig. 5.

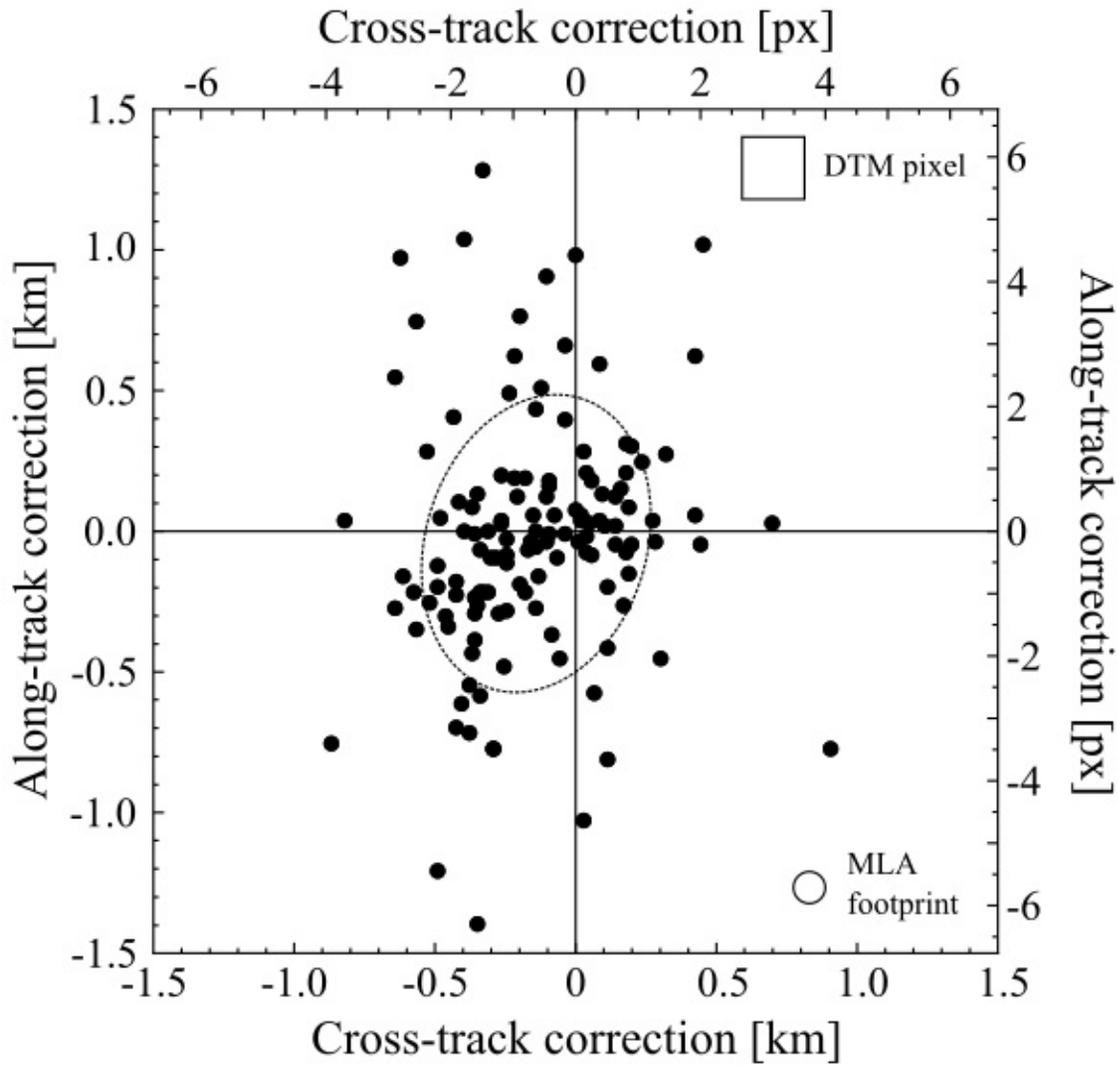

**Fig. 7.** Lateral (along-track and cross-track) corrections for MLA profiles after co-registration with the H6 DTM. The data scatter, which greatly exceeds the co-registration errors (< ½ pixel for most profiles), is thought to be due to spacecraft orbit and attitude uncertainties or residual errors of the Mercury rotation model. The dashed ellipses indicate the 1σ spread of the derived corrections. For comparison the size of a DTM pixel and the median diameter of all MLA footprints are shown

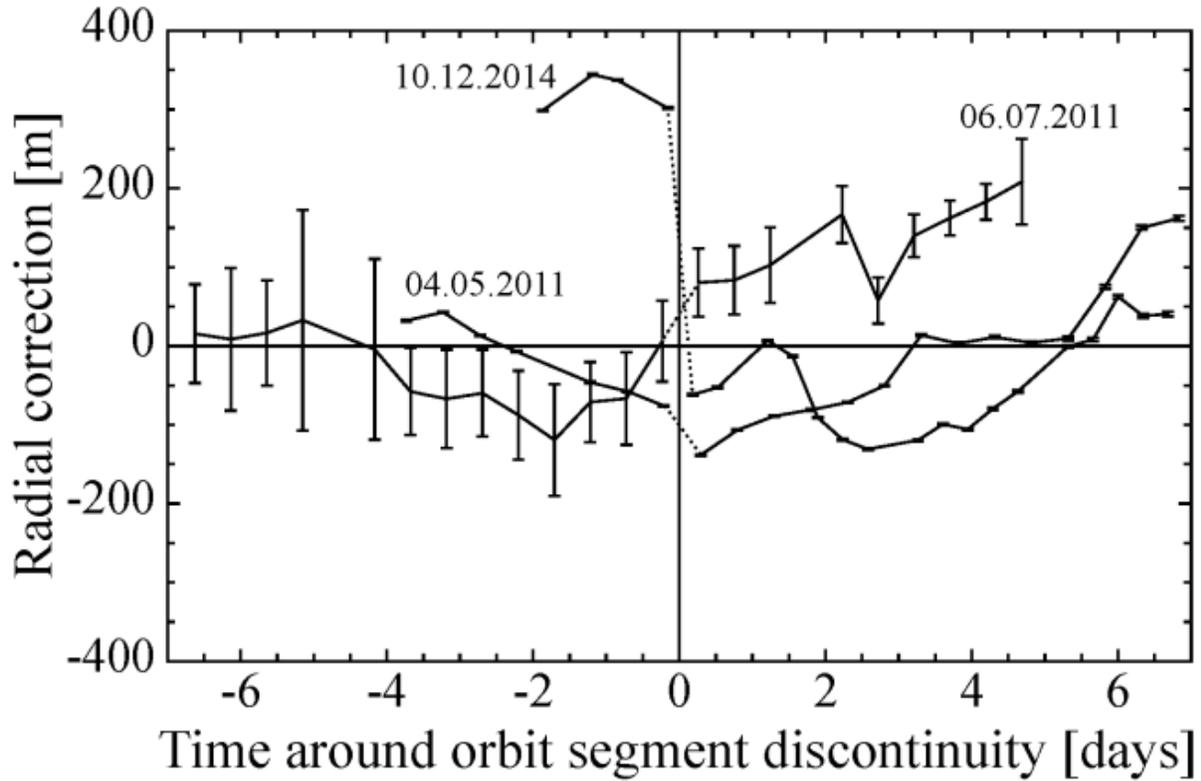

**Fig. 8.** Radial corrections for MLA profiles derived through co-registration to the H6 DTM. The numbers at the curves denote the time of the orbit segment discontinuity event. All curves are centered with respect to their orbit segment discontinuity events. Error bars denote the uncertainty of the radial correction for the vertical co-registration of the respective MLA profiles. See Section 6 for further details.

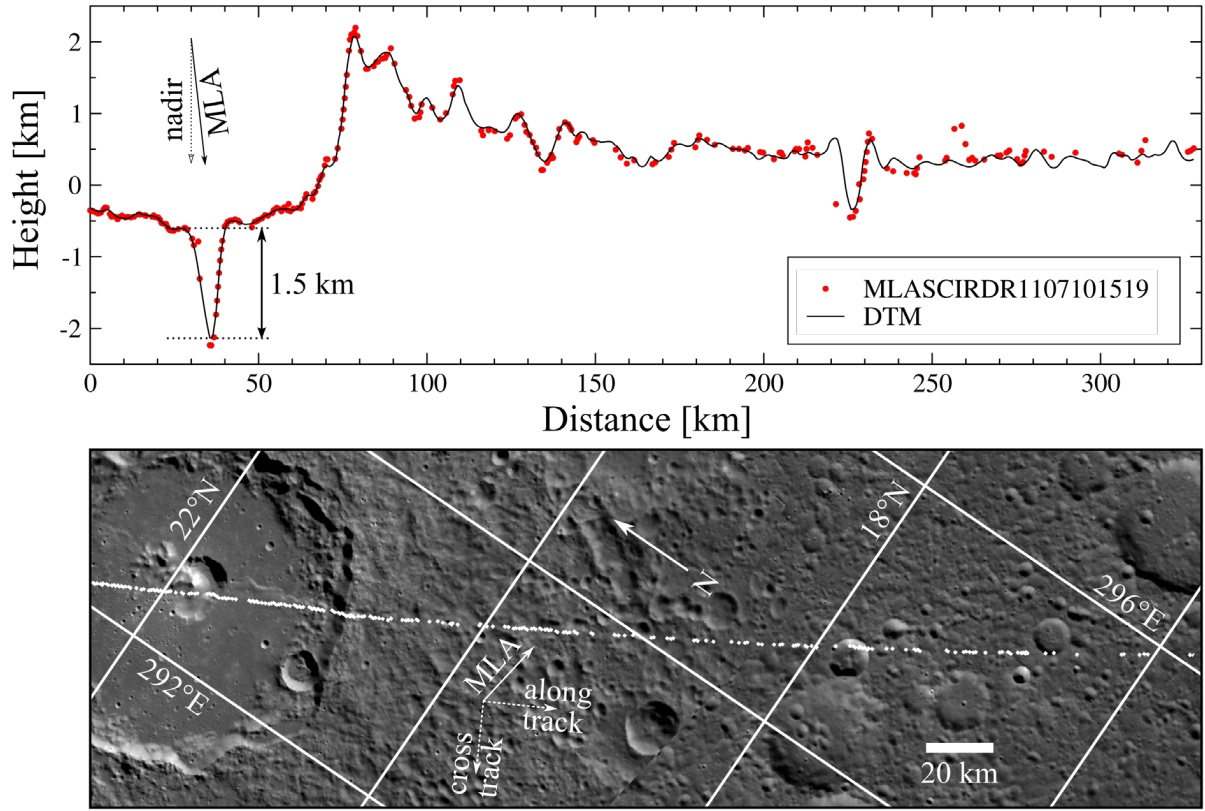

**Fig. 9.** Top: MLA measurements (red dots) and stereo DTM heights (black line). The left part of the track shows a portion of the Catullus crater with its 1.5 km deep pit. Note the gaps and the misidentifications. Bottom: The MLA profile over a MESSENGER MDIS mosaic. The MLA measurements were performed at an off-nadir pointing of 35°.

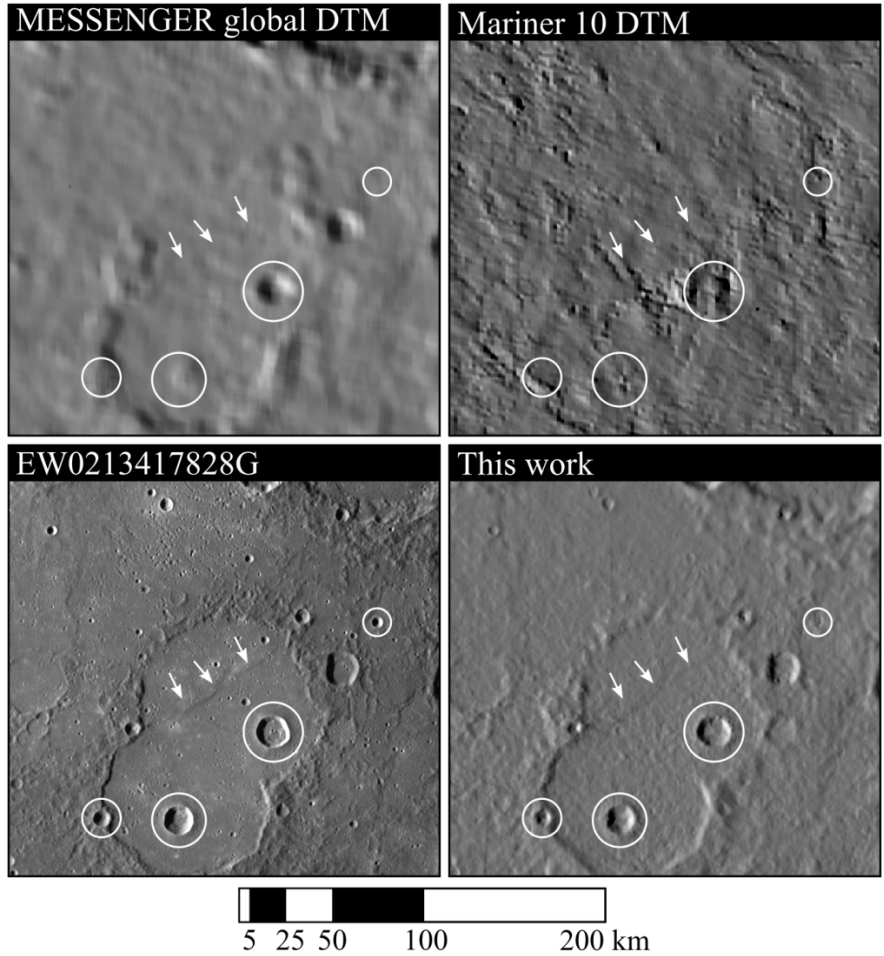

**Fig. 10.** Demonstration of effective resolutions of various DTM products for Mercury. Along with the ortho-rectified MDIS WAC image (subset of EW0213417828G) shaded reliefs of the MESSENGER global DTM (Becker et al., 2016), Mariner 10 DTM (Cook and Robinson, 2000), and of the DTM from this work are shown. All images show the same area on Mercury centered at 7.48°S, 311.75°E. The white circles encompass impact craters with diameters ranging from 6.3 to 17.8 km and the white arrows point to a thrust fault through the lava flooded crater.

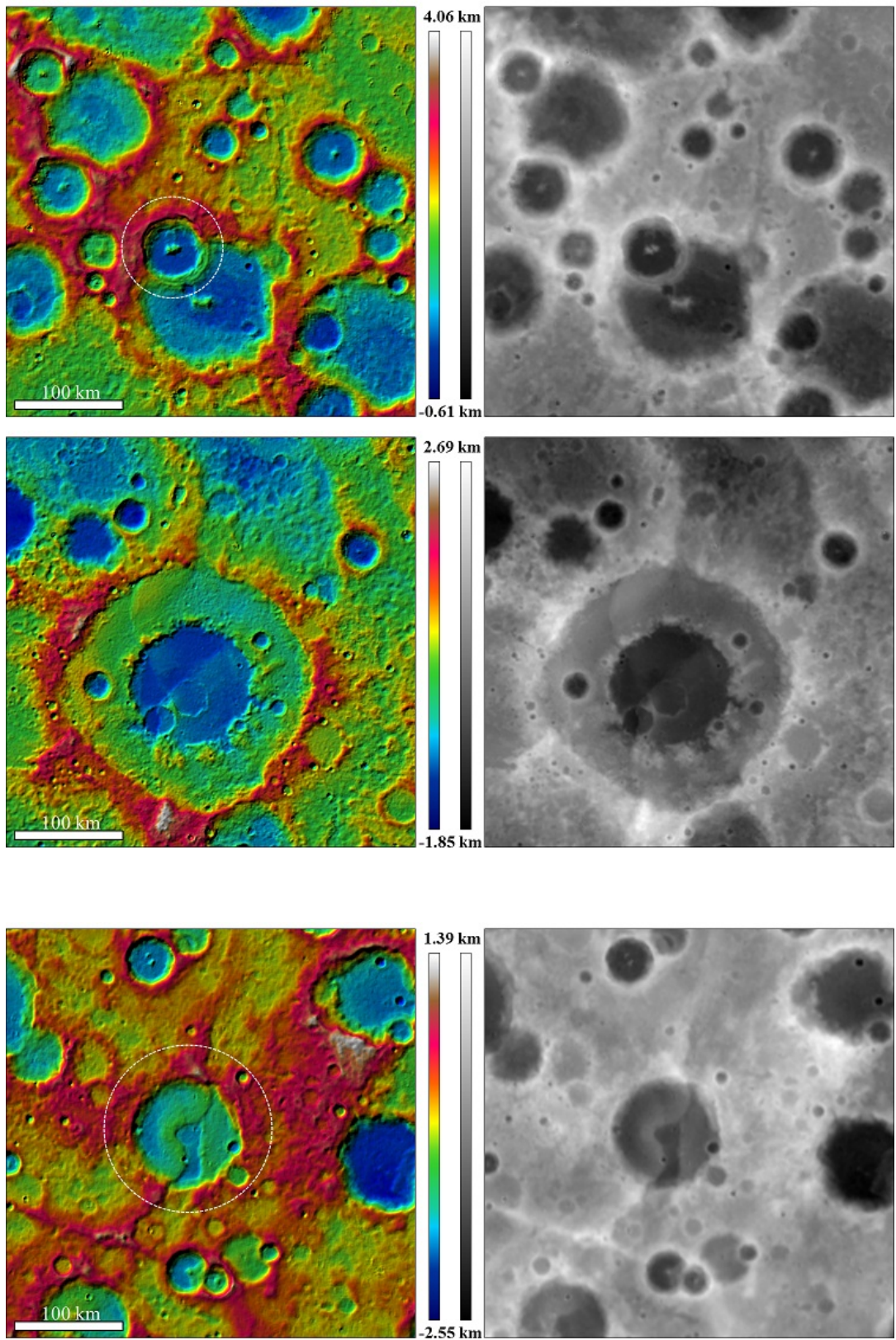

**Fig. 11.** Top panel: The Kuiper crater (11.4°S, 328.8°E) has a diameter of about 60 km diameter and a depth of about 3.1 km. Middle panel: The Renoir basin (18.4°S, 308.2°E) with a diameter of about 240 km shows a well-defined double ring structure, intersected by two scarps. Bottom panel: The Tarkur crater (3.0°S, 295.5°E) has a diameter of about 110 km and a depth of about 2.5 km. Elevation is given w.r.t reference sphere as in Fig. 5.

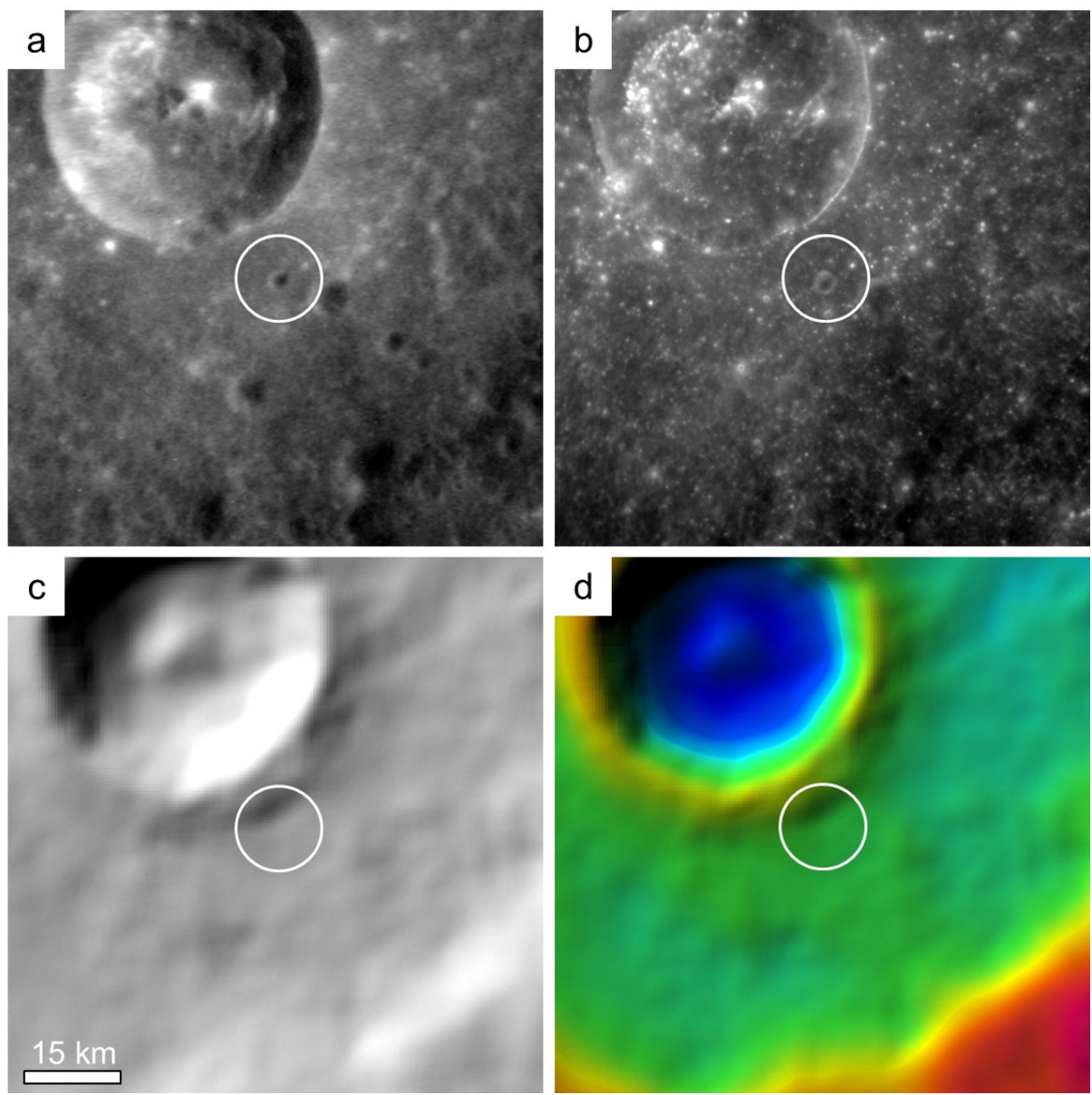

**Fig. 12.** Images (top) (a: EN0131770954M, 126 m/pixel and b: EN1005053163M, 47 m/pixel) ortho-rectified with 50 m/pixel using the H6 DTM and corresponding area of the DTM (bottom, c and d). The crater "Hun Kal", nominal location at 0.5°S, 340°E, highlighted by a white cycle, is seen in the images, but cannot be identified in the DTM.